\documentclass[preprint, superscriptaddress,amsmath, nofootinbib]{revtex4-1}
\usepackage{graphicx}
\usepackage{latexsym}
\usepackage{slashed}
\usepackage{bm}
\usepackage{color}
\usepackage[colorlinks,citecolor=blue,urlcolor=blue,linkcolor=blue]{hyperref}
\usepackage{diagbox}

\usepackage{makecell}

\def\({\left(}
\def\){\right)}

\def\beq{\begin{equation}}
\def\eeq{\end{equation}}

\begin{document}

\title{Deep Learning Jet Image as a Probe of Light Higgsino Dark Matter at the LHC}
	
\author{Huifang Lv}
\email{lvhf@njnu.edu.cn}
\affiliation{Department of Physics and Institute of Theoretical Physics, Nanjing Normal University, Nanjing, 210023, China}

\author{Daohan Wang}
\email{wangdaohan@itp.ac.cn}
\affiliation{CAS Key Laboratory of Theoretical Physics, Institute of Theoretical Physics, Chinese Academy of Sciences, Beijing 100190, China}
\affiliation{School of Physical Sciences, University of Chinese Academy of Sciences, Beijing 100049, China}

\author{Lei Wu}
\email{leiwu@njnu.edu.cn}
\affiliation{Department of Physics and Institute of Theoretical Physics, Nanjing Normal University, Nanjing, 210023, China}

\begin{abstract}
Higgsino in supersymmetric standard models can play the role of dark matter particle. In conjunction with the naturalness criterion, the higgsino mass parameter is expected to be around the electroweak scale. In this work, we explore the potential of probing the nearly degenerate light higgsinos with machine learning at the LHC. By analyzing jet images and other jet substructure information, we use the Convolutional Neural Network(CNN) to enhance the signal significance. We find that our deep learning jet image method can improve the previous result based on the conventional cut-flow by about a factor of two at the High-Luminosity LHC.
\end{abstract}
\maketitle

\tableofcontents

\newpage
\section{Introduction}
\label{sec1}
The existence of dark matter (DM) has been verified in various astrophysics and cosmological observations. One of the attractive candidates for the DM particle is the Weakly-Interacting Massive Particles (WIMPs). When the $R$-parity is conserved, the lightest neutralino $\widetilde{\chi}_{1}^{0}$ in the supersymmetric models can naturally play the role of WIMP dark matter~\cite{Jungman:1995df}. 
When $\widetilde{\chi}_{1}^{0}$ is higgsino-like and light, its annihilation rate is usually too large to saturate the thermal relic density of DM~\cite{Arkani-Hamed:2006wnf}. Thus, other production mechanisms, such as non-thermally produced DM, are needed~\cite{Baer:2014eja}. The phenomenology of light higgsino DM has been studied in Refs.~\cite{Giudice:1995np,Freese:1996kq,Feng:2000gh,Giudice:2010wb,Baer:2011ec,Cao:2012rz,Baer:2013vpa,Kowalska:2013ica,Han:2013kga,Cao:2013mqa,Low:2014cba,Nagata:2014wma,Barducci:2015ffa,Aparicio:2016qqb,Liu:2017msv,Abdughani:2017dqs,Han:2018rkz,Abdughani:2019wss,Baer:2020sgm,Delgado:2020url,Dai:2022isa}.

On the other hand, in the supersymmetric models, the naturalness strongly indicates that the higgsinos mass should not be far above the weak scale~\cite{Feng:1999mn,Feng:1999zg,Papucci:2011wy,Hall:2011aa}. Searching for the light higgsino particles is a key test of the supersymmetric naturalness criteria. 
In MSSM, there are four neutralinos of $\widetilde{\chi}_{1,2,3,4}^{0}$ that are the mixtures of bino $\widetilde{B}$ ,wino $\widetilde{W}$ and neutral higgsinos $\widetilde{H}_{u,d}^{0}$. Besides, there are two charginos of $\widetilde{\chi}_{1,2}^{\pm}$that are the mixtures of wino $\widetilde{W}^{\pm}$ and charged higgsinos $\widetilde{H}_{d}^{-},\widetilde{H}_{u}^{+}$. The neutralino mass matrix is given by
\begin{equation}
\begin{aligned}
&M_{\widetilde{\chi}^{0}} = 
\left( \begin{array}{cccc}
M_{1} & 0 & -M_{Z}s_{W}c_{\beta} & M_{Z}s_{W}s_{\beta} \\
 0 & M_{2} & M_{Z}c_{W}c_{\beta} &  -M_{Z}c_{W}s_{\beta} \\
 -M_{Z}s_{W}c_{\beta} & M_{Z}c_{W}c_{\beta} & 0 & -\mu \\
M_{Z}s_{W}s_{\beta} & -M_{Z}c_{W}s_{\beta} & -\mu & 0 
\end{array}\right),
&M_{\widetilde{\chi}^{\pm}} = 
\left( \begin{array}{cc}
M_{2}& \sqrt{2}M_{W}c_{\beta} \\
\sqrt{2}M_{W}s_{\beta} & \mu
\label{nmatric}
\end{array}\right).
\end{aligned}
\end{equation}
where $M_{1}$ and $M_{2}$ represent the soft breaking mass parameters of bino and wino, respectively. $\mu$ stands for the higgsino mass parameter. $M_{Z}$ and $M_{W}$ are the masses of $Z$ boson and $W$ boson. $s_{W},s_{\beta},c_{W}$ and $c_{\beta}$ represent $\sin{\theta_{W}},\sin{\beta},\cos{\theta_{W}}$ and $\cos{\beta}$, respectively. 

When $\mu \ll M_{1},M_{2}$, the lightest chargino $\widetilde{\chi}_{1}^{\pm}$ and the two lightest neutralinos $\widetilde{\chi}_{1}^{0},\widetilde{\chi}_{2}^{0}$ will be higgsino-like. The strategies of probing these new light particles at colliders are sensitive to the mass difference $\Delta M$ between lightest supersymmetric particle(LSP) and the next-to-lightest supersymmetric particle(NLSP). The conventional methods of searching for the electroweakinos require the hard leptons in the final states. But if $\Delta M$ is small, the decay products of the electroweakinos would become soft. In the past, several new ways of accessing such a compressed spectrum have been proposed at colliders~\cite{Han:2013usa,Arbey:2013iza,Han:2014kaa,Schwaller:2013baa,Han:2015lma,Abdughani:2018wrw,Fukuda:2019kbp,Baer:2021srt}. Among them, the mono-jet signature was used to probe the nearly degenerate higgsinos at the LHC. Due to the large systematics, the additional isolated soft leptons in the final states may help to improve the sensitivity. However, when the mass splitting is small enough, such leptons would fail to reach the lepton triggers in the LHC experiments~\cite{ATLAS:2021moa,CMS:2021edw}.

In recent years, machine learning (ML) has been widely used in the fields of particle collision data processing and classification (see recent reviews, e.g.~\cite{Ren:2017ymm,Guest:2018yhq,Albertsson:2018maf,Abdughani:2019wuv,Feickert:2021ajf,Karagiorgi:2021ngt}). In contrast with the traditional cut-based analysis, ML provides a new way to optimize the signal-to-noise ratios. In Refs.~\cite{Cogan:2014oua,deOliveira:2015xxd}, it is found that the four-momentum information of the constituents inside the jets can be converted into a two-dimensional (2D) image. To deal with the jet images, the convolutional neural network (CNN) and its variants have been often adopted in the analyses~\cite{Aurisano:2016jvx,Komiske:2016rsd,Guo:2018hbv,Lin:2018cin,Kasieczka:2019dbj,Lee:2019cad,Guo:2020vvt,Abdughani:2020xfo,Khosa:2021cyk,Ren:2021prq,Jung:2021tym,Chigusa:2022svv}. For example, in Ref.~\cite{Kasieczka:2019dbj}, they studied the image-based networks to distinguish hadronically decaying top quarks from a background of light quark or gluon jets. In Ref.~\cite{Lee:2019cad}, the different neural network classifiers are trained to distinguish between quark/gluon-initiated jets. By using the multi-channel CNNs, their results can be better than that using the boosted decision tree (BDT). In Ref.~\cite{Khosa:2021cyk}, they use Lund jet plane image data to train CNNs to distinguish hadronically decaying Higgs bosons from the QCD backgrounds. By comparing the cut-based approach with the jet color ring observable, they found that the CNN method can give a higher tagging efficiency for all the cases. In Ref.~\cite{Ren:2021prq}, the authors apply the CNN algorithm to study the photon-jet events from the light axion-like particles and obtain a stronger bound on the axion-photon couplings in the MeV-GeV mass range. 

Given the light higgsinos can not only be served as dark matter but also an important measure of the naturalness, it is meaningful to search for these new particles at the LHC. However, due to the low-momentum SM particles in the final states, it is very challenging to probe these nearly degenerate higgsinos with the conventional analysis method. In this work, we will utilize the deep learning jet image method to search for the light nearly-degenerate higgsino DM through the monojet events in the LHC experiment. As a comparison, we also study the performance of the BDT in our analysis. We note that the leading jets from the initial state radiations in our signal processes $ pp\rightarrow \widetilde{\chi}_{1}^{\pm} \widetilde{\chi}_{1}^{\mp} j, \widetilde{\chi}_{1}^{0} \widetilde{\chi}_{2}^{0} j, \widetilde{\chi}_{1}^{\pm} \widetilde{\chi}_{1,2}^{0} j$ contain more gluon jets than the main backgrounds $Z/W$+jets, while the latter are almost entirely dominated by the quark jets, which have different patterns in the jet images. This feature can be used as a handle to improve signal sensitivity. The paper is organized as follows: in Sec.~\ref{sec2}, we present the simulation of signal and background events and construct the jet image. In Sec.~\ref{sec4}, we describe the neural network architecture in our analysis and discuss the numerical results. Finally, we draw the conclusions in Sec.~\ref{sec6}.

\section{Simulation and Pre-selection}
\label{sec2}

\begin{figure}[h]
\centering
\includegraphics[height=3cm,width=11cm]{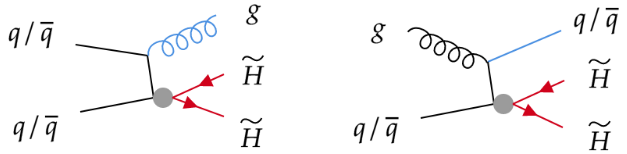}
\caption{Feynman diagrams of our signal processes on the LHC.}
\label{feynman}
\end{figure}
In our simplified nearly-degenerate higgsino scenario, the monojet events are from the processes $ pp\rightarrow \widetilde{\chi}_{1}^{\pm} \widetilde{\chi}_{1}^{\mp} j, \widetilde{\chi}_{1}^{0} \widetilde{\chi}_{2}^{0} j, \widetilde{\chi}_{1}^{\pm} \widetilde{\chi}_{1,2}^{0} j$, whose Feynman diagrams are shown in Fig.~\ref{feynman}. Since $\mu \ll M_{1}, M_{2}$, the production cross sections of the signal events are sensitive to the higgsino mass parameter $\mu$ but almost independent of the value of $\tan\beta$. The dominant SM backgrounds we consider are $Z(\rightarrow \nu \bar{\nu})+{\rm jet}$, $W^{\pm}(\rightarrow l \nu)+{\rm jet}$, and $W^{\pm}(\rightarrow \tau^{\pm} \nu)+{\rm jet}$. We ignore $t \bar{t}$ background because its contribution to the sensitivity is much smaller than others~\cite{Han:2013usa}. In this paper we will explore the potential of using CNN to identify 2D image information of the monojet to distinguish our signal from the background. In addition, we will compare the results obtained from CNN  with that from the BDT method.
%

We use {\textsf SUSY-HIT}~\cite{Djouadi:2006bz} to calculate the mass spectrum of the relevant sparticles. The  parton level signal and backgrounds are generated by {\textsf MadGraph5 aMC@NLO}~\cite{Alwall:2014hca}. The parton shower and hadronization are performed with {\textsf Pythia-8}~\cite{Sjostrand:2014zea}. The detector simulation is carried out by {\textsf Delphes}~\cite{deFavereau:2013fsa}. In Delphes, the electromagnetic calorimeter (ECAL) measures the energy of electrons and photons. All the energies of electrons and photons are deposited in the ECAL by default. The hadron calorimeter (HCAL) measures the energy of long-lived charged and neutral hadrons. Although a fraction of the energy of stable hadrons would be deposited in the ECAL in the realistic detectors, it is assumed that all energy of the hadrons is deposited in the HCAL in the {\textsf Delphes}. Kaons and $\Lambda$ are considered long-lived by most detectors. In {\textsf Delphes}, their energies are assumed to be distributed in the ECAL and HCAL at $f_{{\rm ECAL}}$ = 0.3 and $f_{{\rm HCAL}}$ = 0.7~\cite{deFavereau:2013fsa}. For the reasonable statistics, we imposed a parton-level cut of $p_{T} > 120$ GeV and $|\eta|<5$ on the leading jet when generating the parton-level signal and background events. Finally, with the particle-flow algorithm~\cite{deFavereau:2013fsa,ATLAS:2017ghe,CMS:2017yfk}, the EflowPhotons, the EflowNeutralHadrons, and the ChargedHadrons, consist of the energy deposits in the ECAL and HCAL. They are clustered into jets by {\textsf FastJet}~\cite{Cacciari:2011ma} by using the anti-$k_{t}$ algorithm with $R_{j}=0.7$~\cite{Cacciari:2008gp}. 

\begin{figure}[h]
\centering
\includegraphics[height=7cm,width=8cm]{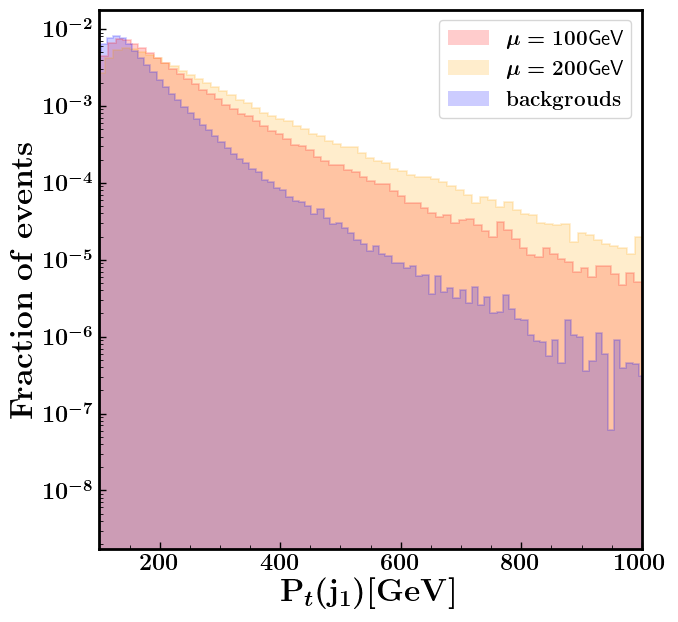}
\includegraphics[height=7cm,width=8cm]{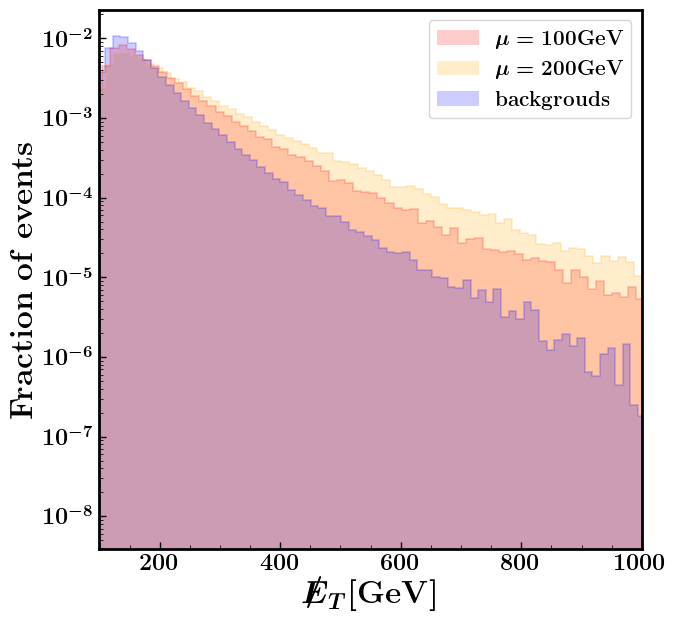}
\caption{The normalized distributions of the leading jet $p_{T}(j_{1})$ and the $\slashed{E}_{T}$ of the signal and background events at the 14 TeV LHC.}
\label{fig2}
\end{figure}

In Fig.~\ref{fig2}, we show the normalized distributions of the leading jet $p_{T}(j_{1})$ and the missing transverse energy $\slashed{E}_{T}$. It can be seen that the transverse momentum of the leading jet in the signal is harder than that in the background, and the $\slashed{E}_{T}$ in the signal drops more slowly than that in the background. Thus, we choose $p_{T}(j_{1}) > 300 $ GeV and $|\eta_{j_{1}}|<2$, and $\slashed{E}_{T} > 300 $ GeV to effectively reduce the background~\cite{Abdughani:2019wss}. For statistics, we also keep the second jet with $p_{T}(j_{2}) < 100$ GeV and $|\eta_{j_{2}}|>2$ in our analysis. At the same time, we veto the events with more than two jets with $p_{T}$ greater than 30 GeV in the area of $|\eta|<4.5$. On the other hand, we veto the events with an identified lepton($l=e,\mu,\tau$) and b-jet to reduce the backgrounds. In addition, we use the requirements of $\Delta\phi(j_{1},\slashed{E}_{T})>0.7$ to reduce most of the QCD backgrounds~\cite{CMS:2018rea,Jorge:2021vpo}. The basic selections of the signal and background events are shown in Tab.~\ref{basiccut}. After the basic cuts, we will train the samples with the BDT and CNN.

\begin{table}[h]
\small
    \centering
    \begin{tabular}{c|c|c|c|c|c}
    \hline
    Basic Cuts & $Z(\nu \bar{\nu})$+j & $W^{\pm}(l \nu_{l})$+j & $W^{\pm}(\tau^{\pm} \nu_{\tau})$+j & signal($\mu$ =100 GeV)&  signal($\mu$ = 200 GeV)\\
    \hline
    \makecell[c]{$p_{T}({\rm j}_{1})>300$ GeV\\$|\eta_{{\rm j}_{1}}|<2$}& 10348 & 9863& 9908& 27544& 43810\\
    \hline
    $\slashed{E}_{T}>300$ GeV&6549 &1804& 3122&19999& 35307\\
    \hline
    \makecell[c]{$p_{T}({\rm j}_{2})<100$ GeV\\$|\eta_{{\rm j}_{1}}|>2$\\$p_{T}({\rm j}_{3})<30$ GeV}&1531 & 385& 478& 3787& 7203\\
    \hline
    veto on e,$\mu, \tau$&1507& 179& 323& 3686& 7080\\
    \hline
    veto on b-jet & 1455& 177& 319& 3558& 6839\\
    \hline
    $\Delta\phi({\rm j}_{1},\slashed{E}_{T})>0.7$&1452& 175& 316& 3536& 6799\\
    \hline
    \end{tabular}
    \caption{The basic cut flows of the cross sections (in units of pb) of the signal and backgrounds at 14 TeV LHC. The higgsino mass $\mu=100,~200$ are taken in the calculations.}    
    \label{basiccut}
\end{table}

\section{Variable Analysis for CNN And BDT}
\label{sec3}

\begin{figure}[h]
\centering
\includegraphics[height=7cm,width=8cm]{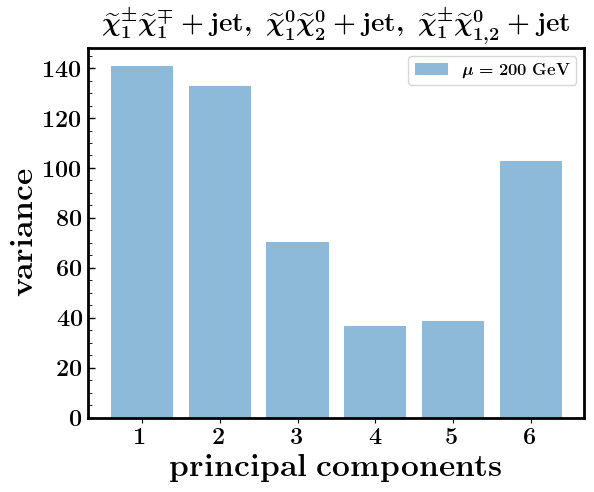}
\includegraphics[height=7cm,width=8cm]{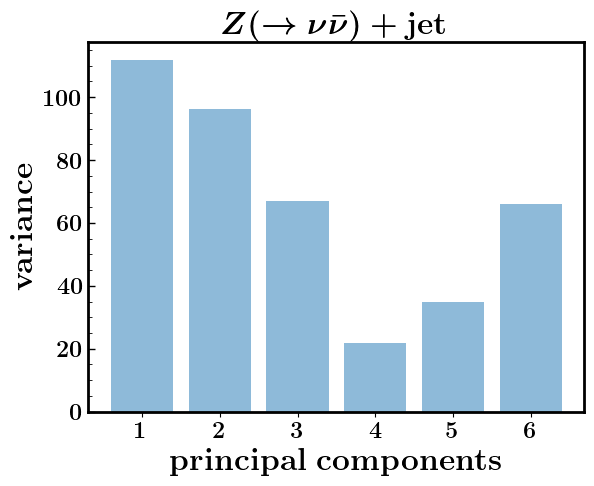}
\includegraphics[height=7cm,width=8cm]{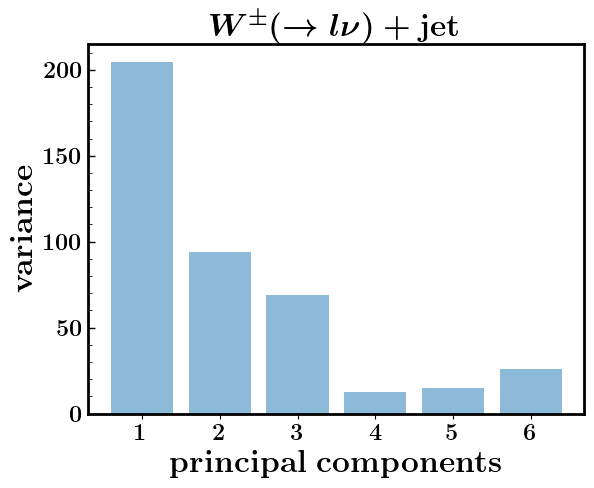}
\includegraphics[height=7cm,width=8cm]{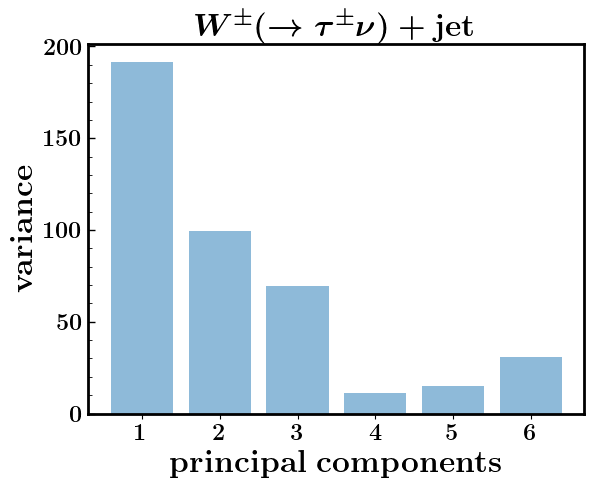}
\caption{The variance of each principal component after principal component analysis (PCA) of the substructure information of monojet in the signal and background. The component with a large variance contains more characteristic information.}
\label{varance}
\end{figure}

\begin{figure}[h]
\centering
\includegraphics[height=7cm,width=8cm]{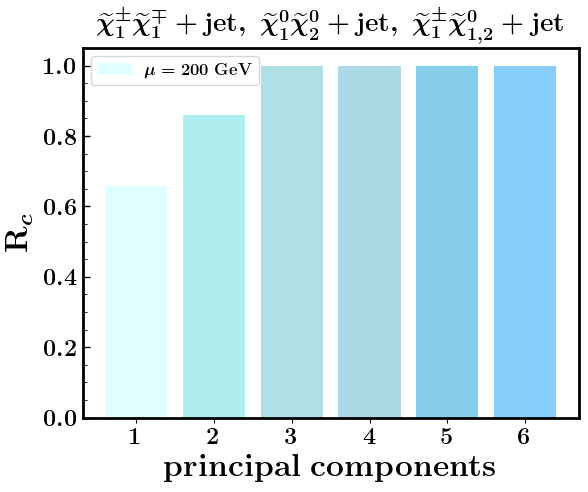}
\includegraphics[height=7cm,width=8cm]{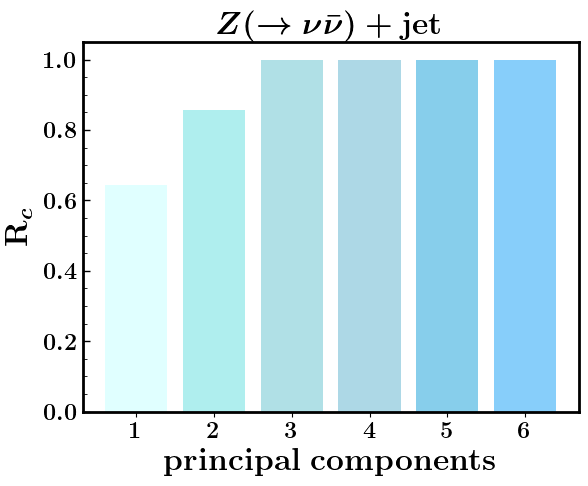}
\includegraphics[height=7cm,width=8cm]{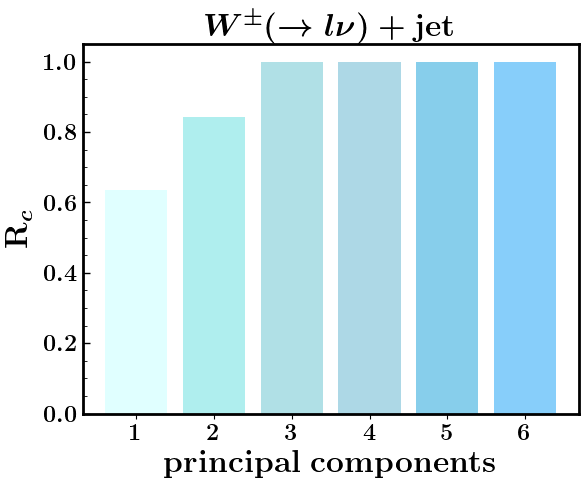}
\includegraphics[height=7cm,width=8cm]{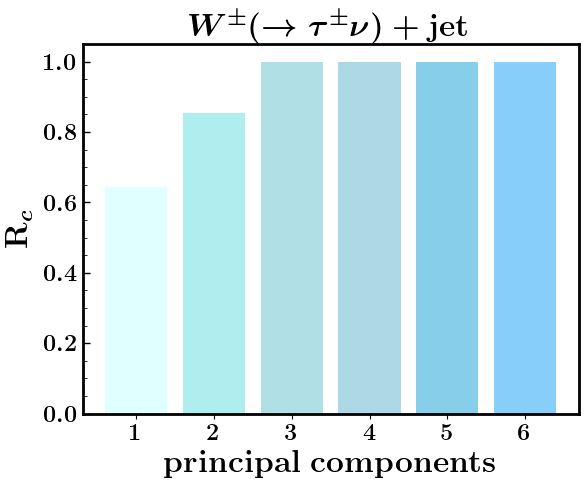}
\caption{The cumulative variance contribution rate (${\rm R}_{c}$) of the principal components. The cumulative variance contribution of six new eigenvariables is equal to one.}
\label{varrate}
\end{figure}

As mentioned above, we used the anti-$k_{t}$ algorithm with $R_{j}=0.7$ to cluster the energy flow objects (composed of deposits in calorimeter cells) into the jet~\cite{Sheff:2020jyw}. The jet substructure usually contain the useful information, such as the transverse momentum $P_{tx}$, the pseudorapidity $\eta$, and the azimuthal angle $\phi$ of each Eflow object, the transverse momentum $P_{tx}^{'}$ of each EflowPhoton, the transverse momentum $P_{tx}^{''}$ of each EflowNeutralHadron, and the transverse momentum $P_{tx}^{'''}$ of each ChargedHadron. We perform the principal component analysis (PCA) on the six-dimensional original variables and plot the histograms of the variance of the new eigenvectors in Fig.~\ref{varance}. In the PCA, a set of possibly correlated variables is transformed into a set of linearly independent variables through the orthogonal transformation, where the transformed set of variables is called the principal components~\cite{Lahav:1994qa,Adanti:1994ua,Khosa:2019kxd,uchiyama2022schrodinger}. In a nutshell, the PCA is a dimensionality reduction method that uses fewer data to represent more salient features. The six new principal component variables given by the PCA are shown in Fig.~\ref{varance}. All these components are linear combinations of the original variables described above. When we use the linear combination of the original variable with maximum variance, the principal components (dimension is $M^{'}$) can effectively represent the characteristics of the original variables (dimension is $M$). In this paper, $M^{'} < M = 6$. From Fig.~\ref{varance}, we can find that the variance of the first principal component is the largest, and represents the direction of the largest variance in the variable.

In Fig.~\ref{varrate}, we plot the distributions of the cumulative variance contribution ratio $R_c$ of the new feature variables for the signal and backgrounds. 

\begin{equation}
R_{c} = \frac{\sum_{i=1}^{n}V_{i}}{\sum_{i=1}^{6}V_{i}}(n=1,2,3,4,5,6),
\end{equation}
where $V_{i}$ represents the variance of the i-th new feature variable, and n represents the top n new variables with the largest variance.
From Fig.~\ref{varrate}, we can find that the values of $R_c$ of the first two principal components have reached about 85\%, while the values of $R_c$ of the first three principal components are almost higher than 99\%. Therefore, two or three principal component variables with large variances are sufficient to represent the variable features of our study.

By comparing the original features and the new principal component axis, we can find out which features play an important role in holding information. We expect the classification performance will be improved by using the principal components as the new feature variables. On the other hand, in our study, to determine features that are most relevant for the 2D analysis, we choose to use the original feature variables, which include the transverse momentum $P_{tx}$, the pseudorapidity $\eta$, and the azimuthal angle $\phi$ of each Eflow object~\cite{Khosa:2019kxd}. At the same time, these three original variables can be easily used to construct the jet images.



To construct the jet images, we use the transverse momentum $P_{tx}$, the pseudorapidity $\eta$ and the azimuthal angle $\phi$ of each Eflow object in the original variables. We integrate the jet composition information provided by ECAL and HCAL into digital images~\cite{deOliveira:2015xxd}. The jet image is a $40\times 40$ square grid, which is centered around the leading jet with the radius $R_{j}=0.7$. The pixel of each grid in the plane of $(\Delta \eta,\Delta \phi)$ corresponds to the sum of the transverse momentum of each particle falling in the grid~\cite{Komiske:2018lor}.  
\begin{figure}[h]
\centering
\includegraphics[height=7cm,width=8cm]{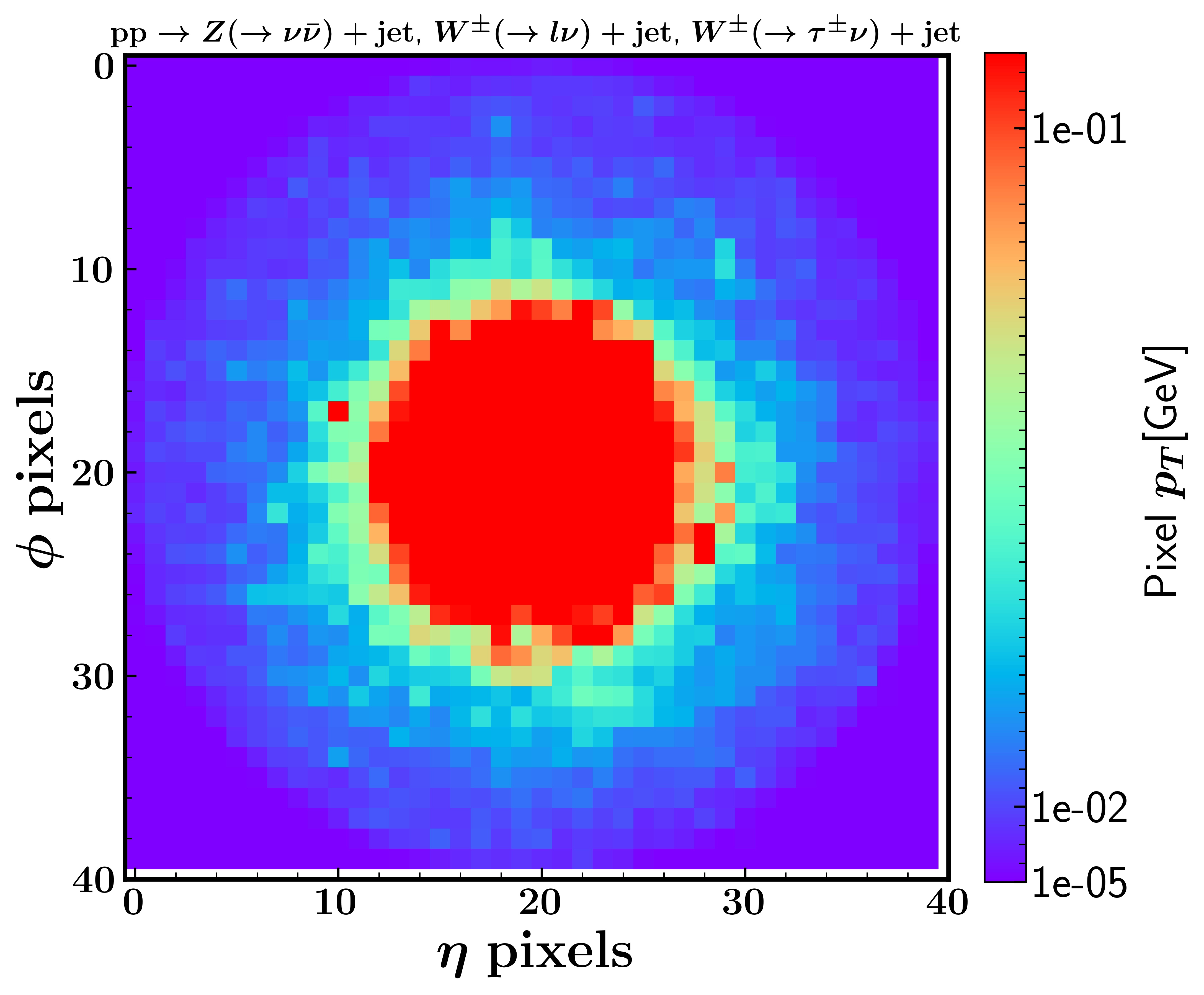}
\includegraphics[height=7cm,width=8cm]{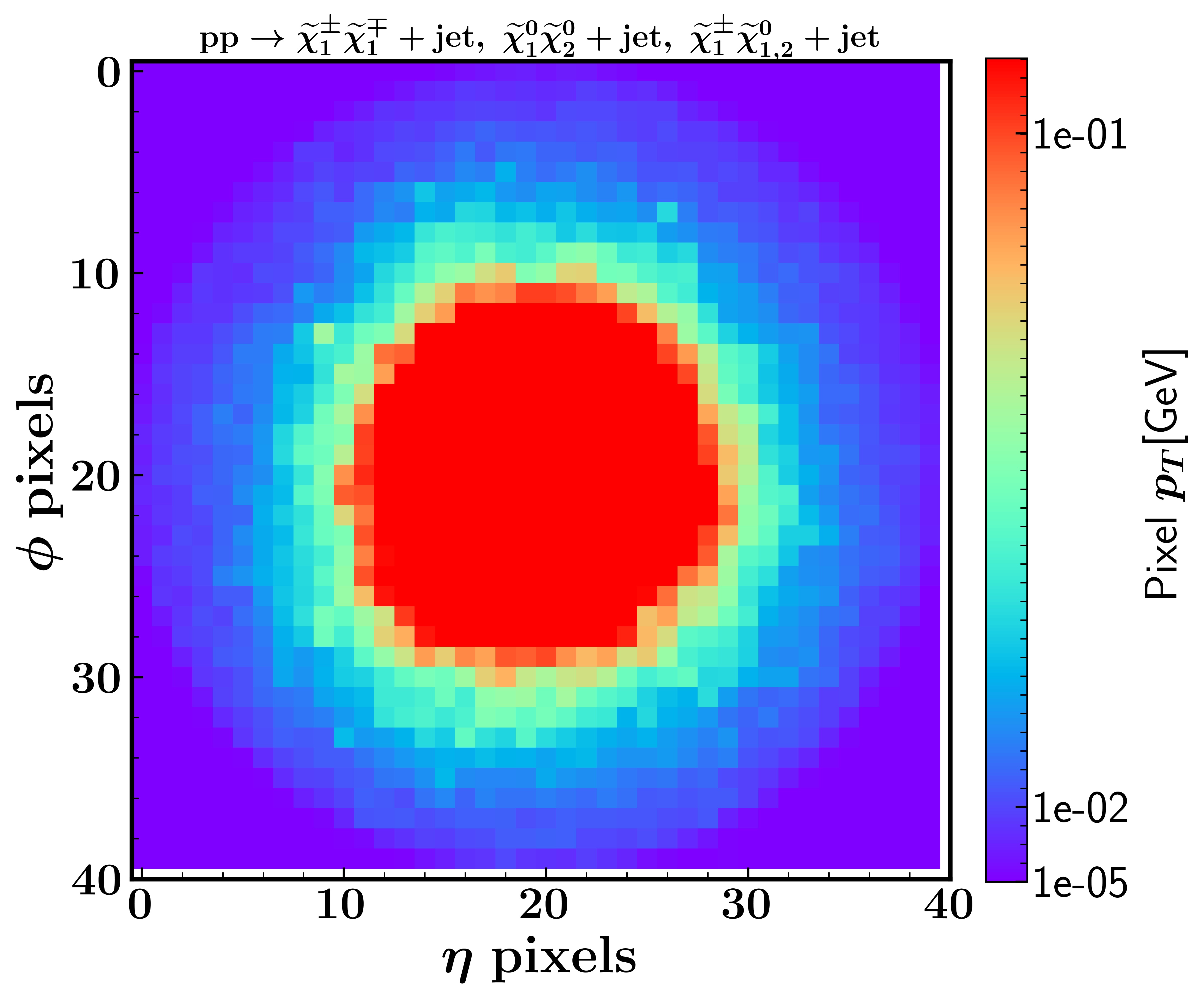}
\caption{The jet images of the leading jet in the total backgrounds (left panel) and signal (right panel) after the translation and rotation. The higgsino mass is taken as $\mu=200$ GeV.}
\label{pixell}
\end{figure}

We pre-process all jet images by using translation, rotation, and normalization. The purpose of pre-processing is to make the CNN learn more about the essential information about the signal and backgrounds. In the translation, we move the center of the jet image $(\eta^{0},\phi^{0})$ to the origin of our new coordinate system $(\eta^{'},\phi^{'})$ as,
\begin{equation}
  \eta^{i\prime}= \eta^{i}-\eta^{0}, \quad \phi^{i\prime} =\phi^{i}-\phi^{0},
\label{translation} 
\end{equation}
where the index $i$ is the number of each pixel. We define the “center of mass” of a jet image as,
\begin{equation}
\eta_{c} = \frac{1}{\sum_{i} p_{T_{i}}} \sum_{i=1}^{n} \eta^{i'}p_{T_{i}}, 
\quad \phi_{c} = \frac{1}{\sum_{i} p_{T_{i}}} \sum_{i=1}^{n} \phi^{i'}p_{T_{i}},
\label{barycenter}
\end{equation}
Then, we can rotate it of each jet image to the vertical axis around its center as,
\begin{equation}
\eta^{i\prime\prime}=\eta^{i\prime}\cos{\alpha}-\phi^{i\prime}\sin{\alpha}, 
\quad \phi^{i\prime\prime}=\phi^{i\prime}\cos{\alpha}+\eta^{i\prime}\sin{\alpha},
\label{rotation}
\end{equation}
with
\begin{equation}
\cos{\alpha}=\frac{\eta_{c}}{\sqrt{\eta_{c}^{2}+\phi_{c}^{2}}}, 
\quad \sin{\alpha}=\frac{\phi_{c}}{\sqrt{\eta_{c}^{2}+\phi_{c}^{2}}}.
\end{equation}

After the translation and rotation, we convert the transverse momentum of the objects into pixel values of the 2D image as, 
\begin{equation}
\begin{aligned} 
&P_{i j}= \sum_{a=1}^{n} P_{a}, \quad (i,j\in[1,40]),\\
&i = \frac{\eta^{i\prime\prime}+0.7}{0.035},\quad
j = \frac{\phi^{i\prime\prime}+0.7}{0.035},
\end{aligned}
\label{calculateptx}
\end{equation}
where $n$ is the number of objects inside the specific grid in each event whose pseudorapidity and azimuthal angle are equal to $(\eta^{i},\phi^{i})$. In Fig.~\ref{pixell}, we show the jet images of the signal and total backgrounds after the translation and rotation, where the pixel value in each grid is the sum average of the transverse momentum $P=P_{tx}$ of the objects inside the jet in each event falling in the corresponding grid. In order to reach higher efficiency in the analysis, we normalize the transverse momentum of the objects within each jet as
\begin{equation}
    y_{i}= \frac{p_{T_{i}}-min(p_{T})}{max(p_{T})-min(p_{T})}.
\label{nomalize}
\end{equation}
where $y_i$ takes the values in the range of [0,1].
\begin{figure}[h]
\centering
\includegraphics[height=7cm,width=8cm]{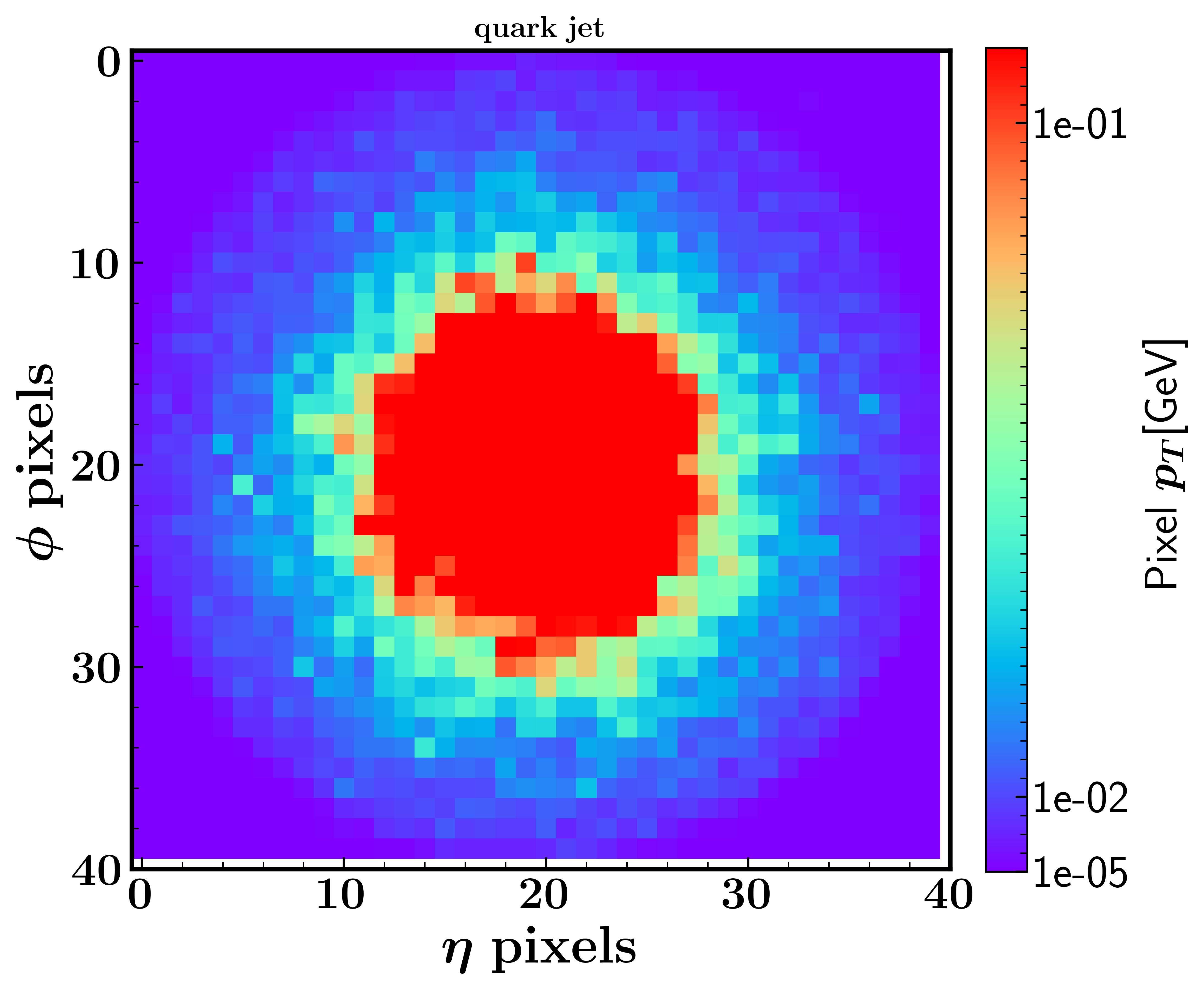}
\includegraphics[height=7cm,width=8cm]{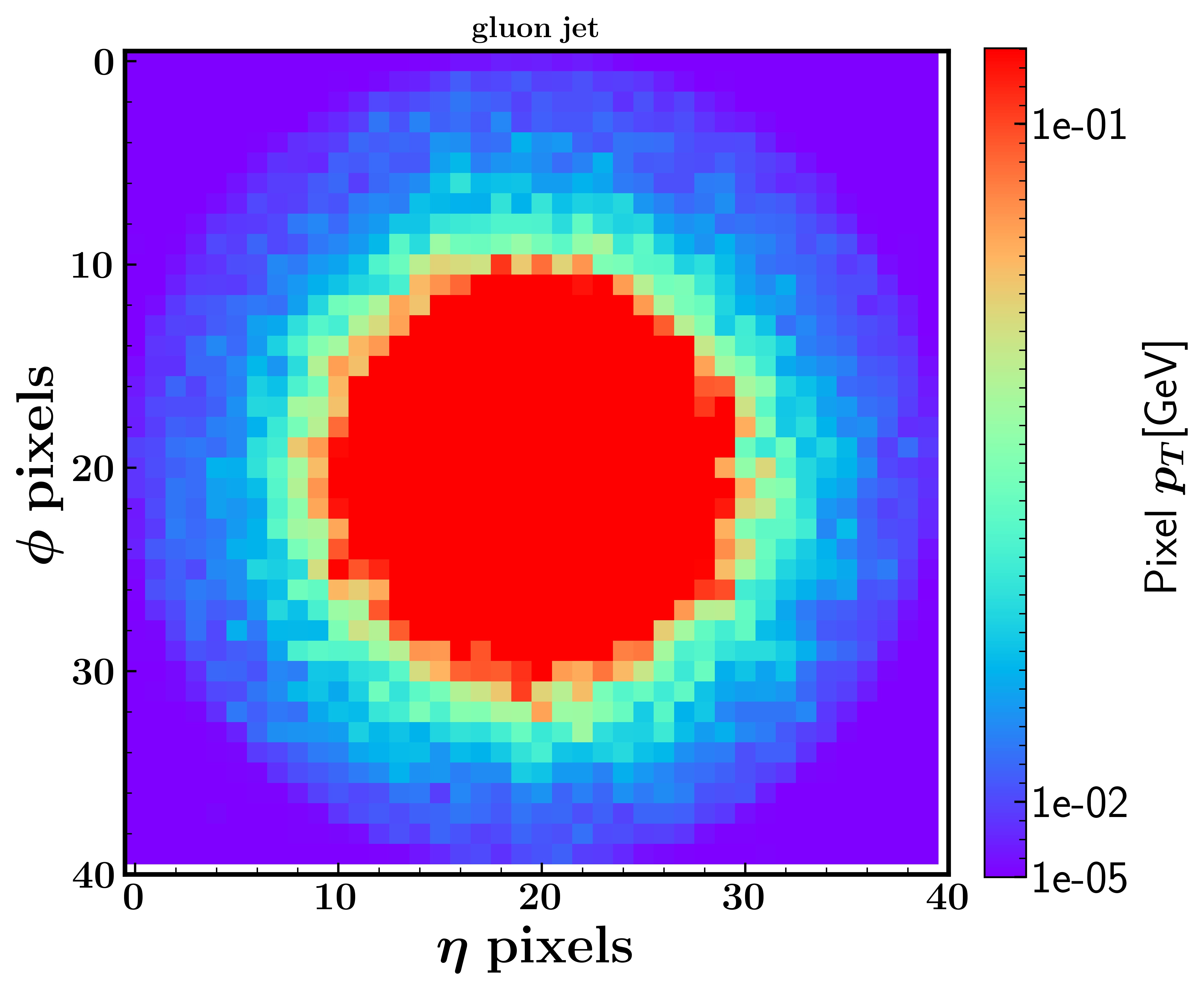}
\caption{The jet images of the quark jet (left panel) and the gluon jet (right panel) in the signal after the translation and rotation. The higgsino mass is taken as $\mu=200$ GeV.}
\label{jet}
\end{figure}

From Fig.~\ref{jet}, we can see that the pixel area of the gluon jets is larger than that of the quark jets. This can be understood that the quark jets carry only one quantum chromodynamic (QCD) color, while the gluon jet has both color and inverse color. Theoretically, the Altarelli-Parisi splitting function~\cite{Altarelli:1977zs} contains a factor $C_{A}=3$ for the gluon radiation from the gluons and a factor $C_{F}=\frac{4}{3}$ for the gluon radiation from the quarks. Therefore, the gluon jets tend to have more composition and broader pixel area in the $\eta$-$\phi$ plane than the quark jets~\cite{ATL-PHYS-PUB-2017-017,Lee:2019cad}. In our calculation, we find that the gluon leading jet in the signal accounts for about 40\%, while its proportion in the background is about 16\%. Thus, the leading jets in the signal events tend to have a broader pixel area than those in the background events.

In order to explore the potential of both CNN and BDT methods to distinguish our signal from the backgrounds and compare their performance, we consider the following jet substructure variables for the implementation of the BDT method according to Refs~\cite{Gallicchio:2012ez,CMS:2021dzg}:

(1) Girth of the jet, which was constructed by adding up the $p_{T}$ deposits within the jet, weighted by distance from jet axis~\cite{Gallicchio:2010sw,Gallicchio:2011xq}.It is defined as $g = \sum_{i \in jet} \frac{p_{T,i} r_{i}}{p_{T}^{jet}}, r_{i} = \sqrt{(\eta_{i}-\eta_{0})^{2}+(\phi_{i}-\phi_{0})^{2}}$,where $p_{T,i}$ is the transverse momentum of the i-th particle inside the jet, $(\eta_{0},\phi_{0})$ is the jet coordinates in the frame with the interaction point of p-p collision as the origin. 

(2) Fragmentation distribution of the jet, which is defined as $p_{T}D=\frac{\sqrt{\sum_{i}p_{T,i}^{2}}}{\sum_{i}p_{T,i}}$~\cite{CMS:2013kfa,Lee:2019ssx}. 

(3) Shape of the jet, which can be approximated by an ellipse that has two principal axes, the major and the minor axis, and the orientation of the major axis in the $(\eta, \phi)$ plane. We can describe it with a $2\times2$ matrix $M$ consisting of the following elements: $M_{11}=\sum_{i}p_{T,i}^{2}\Delta\eta_{i}^{2}$, $M_{22}=\sum_{i}p_{T,i}^{2}\Delta\phi_{i}^{2}$, $M_{12}=M_{21}=-\sum_{i}p_{T,i}^{2}\Delta\eta_{i}\Delta\phi_{i}$, $(\Delta\eta=\sqrt(\eta_{i}-\eta_{0})^{2},\Delta\phi=(\phi_{i}-\phi_{0})^{2})$. The major $(\sigma_{1})$ and minor $(\sigma_{2})$ axes of the jet can be computed from the eigenvalues $\lambda_{1,2}$ of the matrix $M$ by: $\sigma_{1}=\sqrt{\frac{\lambda_{1}}{\sum_{i}p_{T,i}^{2}}}$, $\sigma_{2}=\sqrt{\frac{\lambda_{1}}{\sum_{i}p_{T,i}^{2}}}$~\cite{CMS:2013kfa}.

(4) Distributions of jet mass, which is defined as $m_{j}=\frac{m_{jet}}{p_{T}^{jet}}$. The mass of the jet measures how to spread out the constituents of the jet are~\cite{Gallicchio:2012ez}.

(5) The number of constituent particles $n$ in a given jet~\cite{CMS:2013kfa}.

(6) The angle between the jet and the missing transverse momentum, $\Delta\phi(j_{1},\slashed{E}_{T})$~\cite{CMS:2021dzg}.

(7) The energy of the monojet, $E_{jet}$~\cite{CMS:2021dzg}. 
 
(8) N - subjettiness, which can be defined as $\tau_{N}=\frac{\sum_{i}p_{T,i}min \left\{\Delta R_{1,i},\Delta R_{2,i},...,\Delta R_{N,i}\right\}}{\sum_{i}p_{T,i}R_{0}}$, where $k$ runs over all objects in a given jet, $\Delta R_{J,i}=\sqrt{(\Delta\eta)^2+(\Delta\phi)^2}$ is the distance in the rapidity-azimuth plane between a candidate subjet $J$ and a constituent particle $k$, and $R_{0}$ is the characteristic jet radius used in the original jet clustering algorithm~\cite{Thaler:2010tr,Wang:2021uyb}. In this paper, the variables we use are $\tau_{21}=\frac{\tau_{2}}{\tau_{1}}$, $\tau_{31}=\frac{\tau_{3}}{\tau_{1}}$ and $\tau_{32}=\frac{\tau_{3}}{\tau_{2}}$.

The first three variables are widely used to distinguish between quark/gluon jets, with the gluon jets exhibiting larger values of $g_{jet}$, $\sigma_{1}$ and $\sigma_{2}$, and smaller values of $p_{T}D$[78]. Considering that our signal contains more gluon jets than the background, these variables will help to improve the discrimination efficiency. However, in order to improve the discrimination efficiency of BDT as much as possible, we also consider the following five jet substructure variables.

Regarding the base cut, unlike CNN, we chose $p_{T}(j_{1}) > 500 $ GeV and $\slashed{E}_{T} > 500 $ GeV in order to combine BDT with traditional cut flow methods for better signal saliency. Then we use the ROOT package TMVA~\cite{Hocker:2007ht} and the ``BDTD'' option to train our BDT classifier. We use an ensemble of 180 trees with a minimum training node requirement of 2.5\% in each node, a maximum tree depth of 2 and other parameters are set to default values. We use 10k signal events and 15k background events for BDT training and testing with the same number of events. Finally, to avoid the risk of overfitting, we require the Kolmogorov-Smirnov test of the BDT analysis to be greater than 0.01~\cite{Wang:2021uyb}.

\section{Neural Network Architecture and Numerical Results}
\label{sec4}
We design a convolutional neural network (CNN) within the deep learning framework of PyTorch~\cite{2019arXiv191201703P}. We use the GPU to accelerate the training of the classifier. In the CNN, we use the convolution kernel to extract the features of the input image and the activation function to increase the expression ability of the model. Then we employ the maximum pooling layer to reduce the dimensionality of the network to accelerate the training process. After that, we use the gradient descent method to find the minimum loss function value, and finally, adopt the fully connected layer for classification. The first three steps above realize the mapping of the original data to the hidden layer feature space, and the last step is to map the learned feature distribution to the sample label space for classification prediction. 

\begin{figure}[h]
\centering
\includegraphics[height=6cm,width=16cm]{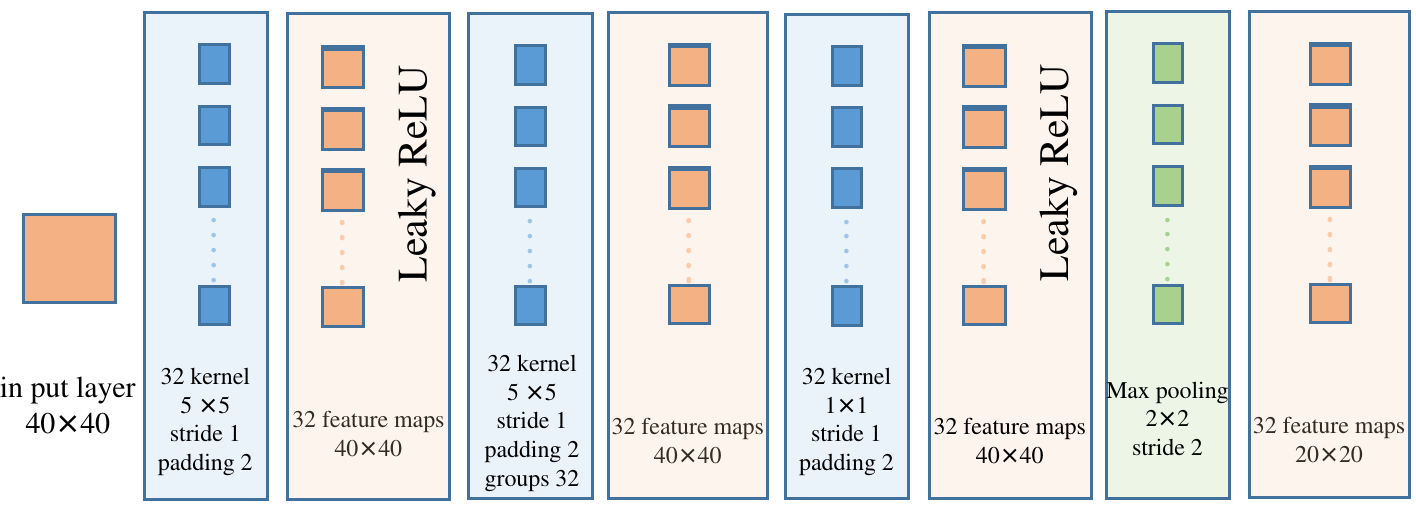}
\includegraphics[height=6cm,width=16cm]{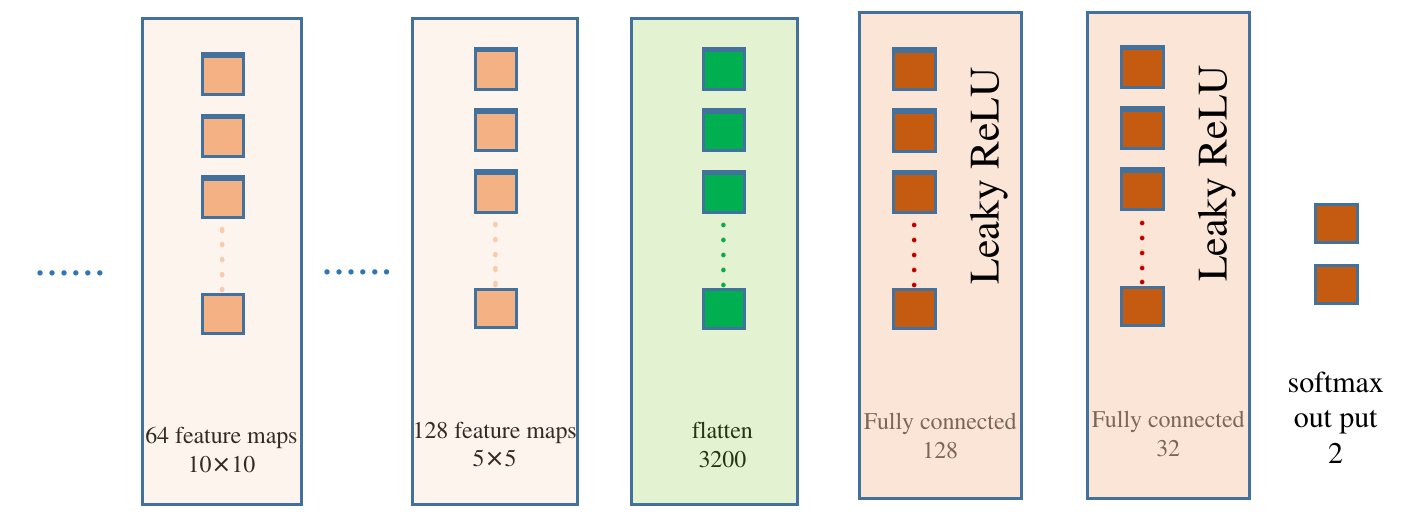}
\caption{The architecture of our convolutional neural network.}
\label{architecture}
\end{figure}

As shown in Fig.~\ref{architecture}, the input 2D image pixels are $40\times 40$ with one channel. The number of convolution kernels is 32, the size is $5\times 5$, and the stride is two. Before the convolution operation, we add two layers of zero at the periphery of the input image to fully extract the edge information of the image and make the image maintains the same dimensions after convolution. Then, we use a non-linear activation function to increase the expressive power of the model. Since both the Sigmoid-type function and the tanh(x)-type function have gradient saturation effects, and the ReLU function has a ``dead zone'', the above three methods are not conducive to gradient convergence. Thus, we choose the Leaky ReLU function here and find that it is indeed better than the above three activation functions. In order to reduce the number of parameters and speed up the operation, we use grouped convolution so that the convolutional kernels in each group are convoluted with only one feature map. 

In order to decrease the sensitivity of the network to the absolute location of elements in the image and reduce parameters, we apply a stack of $2\times 2$ maxpooling layers to obtain the $20\times20$ feature maps with 32 channels. Next, we repeat the above process to obtain the $5\times5$ feature maps of 128 channels. Whereafter, we flatten the final feature map to a single vector and apply two fully connected layers with 128 and 32 neurons respectively. Finally, we apply the softmax activation function to output the probabilities of the signal and background.
\begin{equation}
    s=\frac{e^{S_{i}}}{\sum_{j=1}^{C} e^{S_{ij}}} (i=1,2,3,......N),
\label{softmax}
\end{equation}
where $N$ is the total number of the samples. $C$ refers to the number of categories and is taken as $C$=2 in our study. The softmax function $s$ allows the output values of a multiclass to be converted into a probability distribution with a sum of one in the range of [0, 1]. The network architecture we use is shown in the Fig.~\ref{architecture}.

In addition to the above feedforward operation, a complete machine learning network also needs feedback operation to adjust network parameters and optimize the classification accuracy of the model. In our feedback operation, we define the cross-entropy loss function ${\cal L}$ to calculate the loss rate, 
\begin{equation}
    \mathcal{L}=\frac{1}{N}\sum_{i=1}^{N}(y_{i0}\log{p_{i0}}+y_{i1}\log{p_{i1}}).
\label{cross_entropy loss}
\end{equation}
where $N$ is the number of samples, $y_{{i0},{i1}}$ and $p_{{i0},{i1}}$ are the true and predicted category of the signal and background samples, respectively. We take the Adam optimizer with a learning rate of 0.001~\cite{Kingma:2014vow}.

 
The data set of training and validation contains 400k events, respectively. During the training, the minimum batch of input data is 1024 event samples, and the maximum epoch is set to 100. After the end of the first epoch, each event in the training set is used once. In order to avoid learning the distribution order characteristics of the cases, we randomly scramble the order of the data before each new epoch starts. We also use the same amount of signal and background events as the validation set. We find that the validation set loss function value reaches the minimum as the training reaches about the 43rd epoch. After this epoch, the classifier will learn more detailed features in the training data and then results in over-fitting. Since then, the recognition efficiency of the data in the training set became higher and higher, however, the recognition efficiency of the new batch of validation set data became worse and worse. We stop the training at the 43rd epoch. Therefore, we choose the classifier with the smallest loss function value in the validation data set as our best classifier.

\begin{figure}[h]
\centering
\includegraphics[height=8cm,width=8cm]{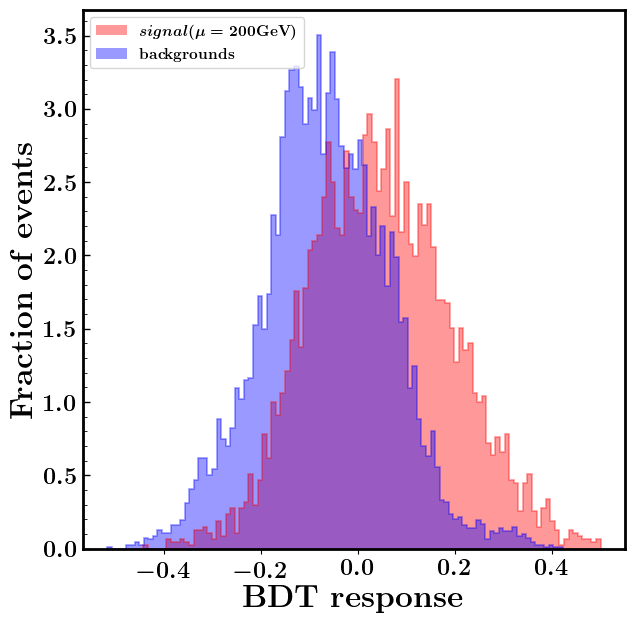}
\includegraphics[height=8cm,width=8cm]{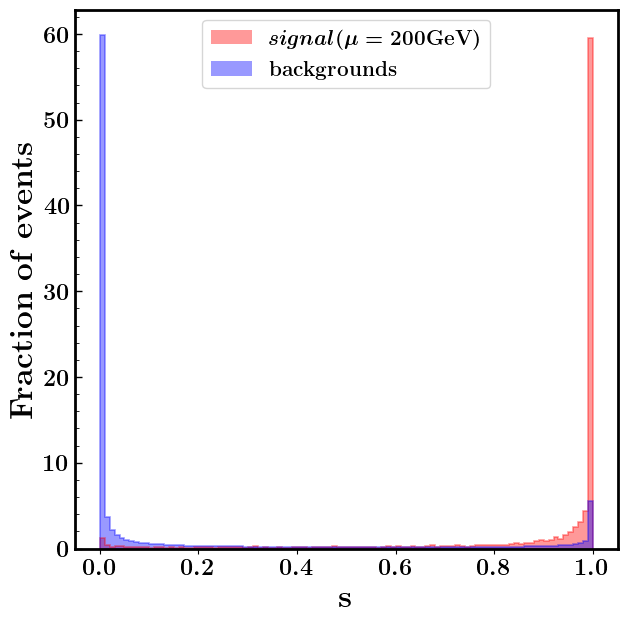}
\caption{The BDT response distributions diagram of the test set of signals and backgrounds (left panel); The CNN classification effect diagram of the test set of signals and backgrounds (right panel).}
\label{loss}
\end{figure}

In the left panel of Fig.~\ref{loss}, we show the BDT responses of signal and background. We can see that the signal events tend to get a higher BDT response, while the background events tend to get a lower response. In the right panel of Fig.~\ref{loss}, we show the test set classification effect for the signal and background events after training the CNN, where $s$ denotes the probability that a sample will be determined as a signal sample after passing through the classifier. The area under the red and blue lines represents the probability density distribution of the signal and background samples, respectively. The former is very close to one, while the latter is very close to zero. However, the irreducible background $pp\rightarrow Z(\rightarrow \nu \bar{\nu})+{\rm jet}$ can contribute a small portion in the region of $s\sim 1$. By comparison, we can find that the discrimination efficiency of BDT is far worse than that of CNN.

\begin{figure}[h]
\centering
\includegraphics[height=8cm,width=8cm]{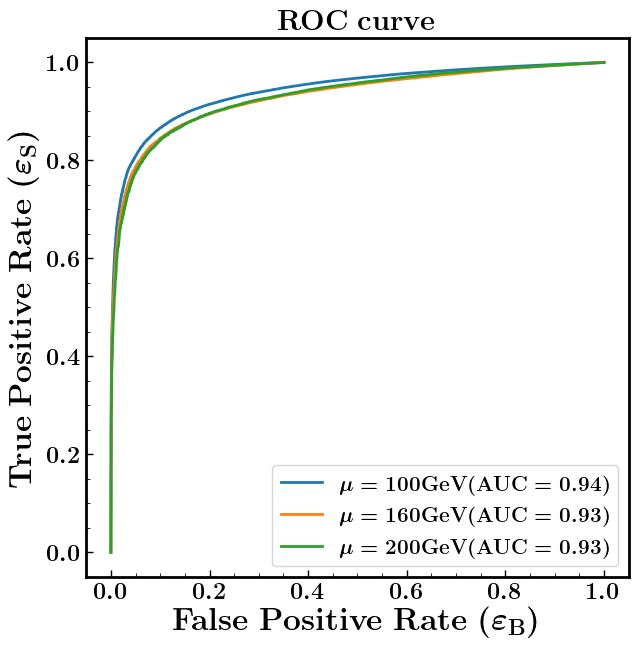}
\includegraphics[height=8cm,width=8cm]{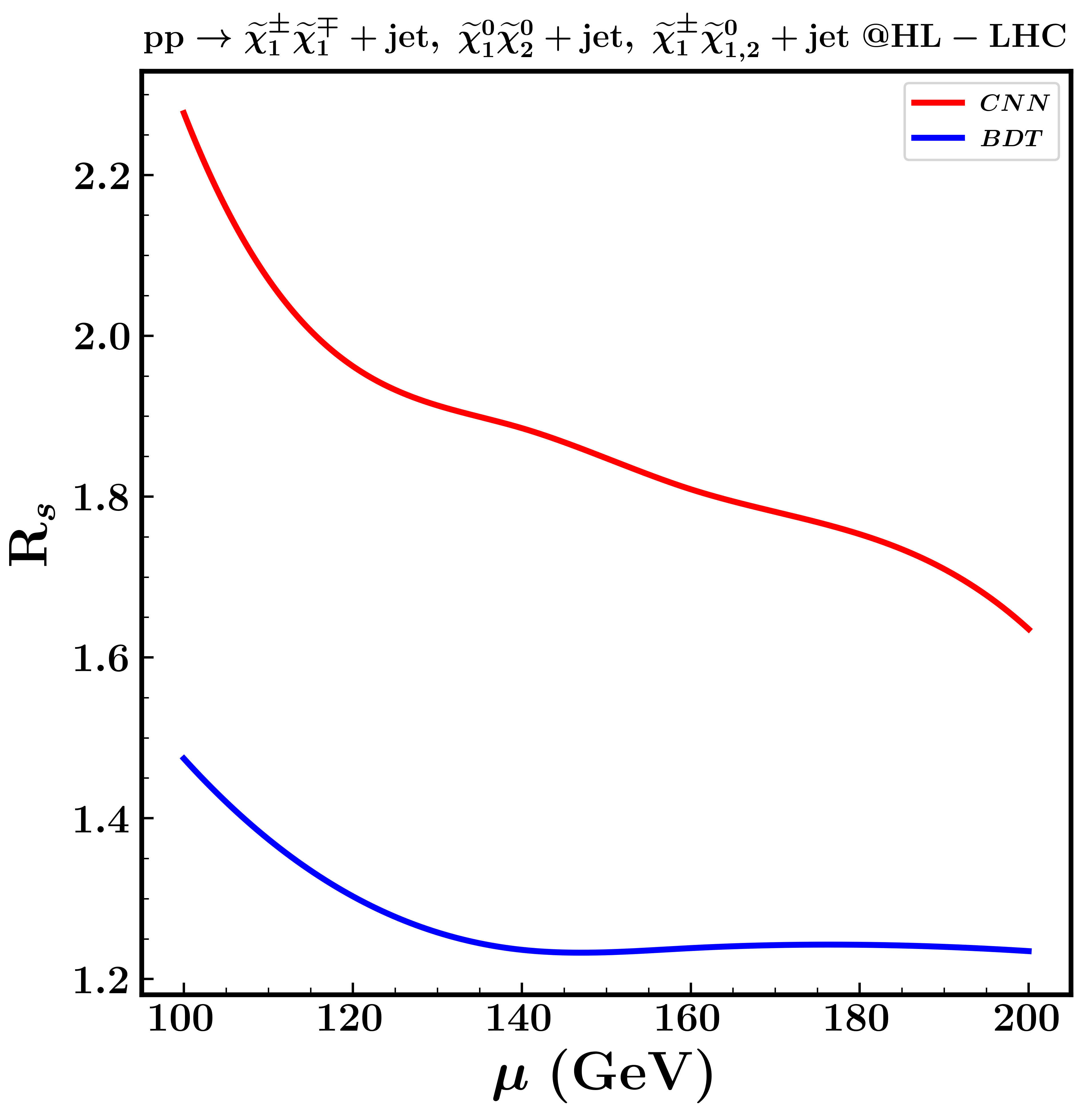}
\caption{The ROC curves and the corresponding AUC values of CNN (left panel). The significance ratio $R_s$ of the CNN (red line) and the BDT (blue line) results $Z_{cnn}$ to the cut-based result $Z_{cut-based}$ (right panel) at the HL-LHC.}
\label{roc}
\end{figure}

In the left panel of Fig.~\ref{roc}, we show the receiver operating characteristic (ROC) curves and the corresponding values of the area under the curve (AUC), where $\mu=100,160,200$ GeV. The true positive rate ($\varepsilon_{S}$) and false positive rate ($\varepsilon_{B}$) represent the fraction of the survival events in the initial signal and background events, respectively. The ROC curves can be obtained from the probability density distribution in the right panel of Fig.~\ref{loss}. If choosing the cut value $s_{0}$, we can use the respective probability densities of the signal and background in the $s > s_{0}$ region as the function values in the ROC curve. When the AUC value approaches one, the classification effect of the classifier is better.

To obtain the observability of the signal, we assume the Poisson distribution to calculate the signal significance $Z$,
\begin{equation}
Z=\frac{S}{\sqrt{B+(\beta B)^2}},
\label{cross_founction}
\end{equation}
where $S$ and $B$ denote the number of the signal and background events after all cuts, respectively. We require the signal events larger than 20 to ensure the statistics and estimate the contribution of the systematic error as $\beta B$. In comparison with the previous results based on the cut-flow in Ref.~\cite{Han:2013usa}, we also focus on the same mass range of $[100,200]$ GeV and calculate the significance with the same value of $\beta=1\%$ at 14 TeV LHC with the integrated luminosity of 3000 fb$^{-1}$. In addition, the basic cut of $p_{T}(j_{1})$ and $\slashed{E}_{T}$ in the Ref.[29] and the analysis based on BDT method in this paper are $p_{T}(j_{1}) > 500 $ GeV and $\slashed{E}_{T} > 500 $ GeV, but since CNN method requires a huge number of events and the computer power is limited, only $p_{T}(j_{1}) > 300 $ GeV and $\slashed{E}_{T} > 300 $ GeV cuts are made in the analysis based on CNN in this paper. To show the improvement, we define the significance ratio of the CNN result $Z_{cnn}$ (red line) or the BDT result $Z_{bdt}$ (blue line) to the cut-based result $Z_{cut-based}$ as,
\begin{equation}
    R_{s} = \frac{Z_{cnn(bdt)}}{Z_{cut-based}}
\end{equation}
From the right panel of Fig.~\ref{roc}, we can see that the CNN method outperforms the BDT method, and the image recognition technology using the CNN can greatly improve the traditional cut-based significance by a factor of two in the low mass range. As the higgsino mass increases, the improvement will decrease because the production cross section of the signal process becomes small. Therefore, when the light higgsinos are highly-degenerate and the soft leptons cannot be used as triggers, the monojet signal will play an important role in searching for these light higgsinos. The application of deep learning jet image may be able to effectively enhance the sensitivity of probing the light nearly-degenerate higgsinos from the mono-jet events at the HL-LHC. 

Besides CNN and BDT used in our paper, there are other more advanced machine learning methods that can be applied to distinguish our signal and backgrounds. For example, Graph Auto-encoder (GNN based on autoencoder) which uses the jet image as input data can be used to distinguish QCD jets and non-QCD jets~\cite{2016arXiv161107308K,Atkinson:2021nlt}; LundNet based on Graph Neural Network, which uses jet on the Lund plane as input data, can be used for W tagging, top tagging, and gluon/quark jet discrimination ~\cite{Dreyer:2018nbf,Dreyer:2020brq}; ParticleNet based on Dynamic Graph Convolutional Neural Network which operates directly on particle clouds for jet tagging, can be used to distinguish gluon jets and quark jets~\cite{Qu:2019gqs}; Attention Based Neural Network can be used to distinguish the jets initiated by the desired process from the general and overwhelming QCD jet and other types of networks~\cite{Li:2020bvf}. These neural networks may be useful for improving the sensitivity of our signal, which deserves further studies in the future.

\section{Conclusion}   
\label{sec6}
In this paper, we study the search of the light higgsino dark matter within the natural MSSM at $14 $ TeV-LHC with $3000\ fb^{-1}$ luminosity. The production processes of our signal are mainly $ pp\rightarrow \widetilde{\chi}_{1}^{\pm} \widetilde{\chi}_{1}^{\mp}+{\rm jet}, \widetilde{\chi}_{1}^{0} \widetilde{\chi}_{2}^{0}+{\rm jet}, \widetilde{\chi}_{1}^{\pm} \widetilde{\chi}_{1,2}^{0}+{\rm jet}$. We convert the information of a jet into a 2D jet image and use the CNN to deeply learn the signal and background image features to enhance the signal observability. In addition, we also use the jet substructure variables and BDT method to distinguish the signal from backgrounds. Our numerical calculation shows that the sensitivity based on CNN can be about $[1.6,2.2]$ times as large as that based on the cut-flow method.
Apart from this, the sensitivity based on BDT is also higher than that based on the traditional cut flow method, but is lower than that based on CNN. The bound obtained from the CNN is about [1.3, 1.5] times that from the BDT method.

\section{Acknowledgements}
We thank Yuchao Gu, Song Li, and Jie Ren for their helpful discussions. In addition, we would like to thank Prof. Jin Min Yang for his warm hospitality at the Institute of Theoretical Physics, Chinese Academy of Sciences.  


\bibliography{refs}

\end{document}